\def\MPL #1 #2 #3 {Mod.~Phys.~Lett.~{\bf#1},\  #2 (#3)}
\def\NPB #1 #2 #3 {Nucl.~Phys.~{\bf#1},\  #2 (#3)}
\def\PLB #1 #2 #3 {Phys.~Lett.~{\bf#1},\  #2 (#3)}
\def\PR #1 #2 #3 {Phys.~Rep.~{\bf#1},\ #2 (#3)}
\def\PRD #1 #2 #3 {Phys.~Rev.~{\bf#1},\  #2 (#3)}
\def\PRL #1 #2 #3 {Phys.~Rev.~Lett.~{\bf#1},\  #2 (#3)}
\def\RMP #1 #2 #3 {Rev.~Mod.~Phys.~{\bf#1},\  #2 (#3)}
\def\ZP #1 #2 #3 {Z.~Phys.~{\bf#1},\  #2 (#3)}
\def\IJMP #1 #2 #3 {Int.~J.~Mod.~Phys.~{\bf#1},\  #2 (#3)}

\def\br{B}
\def\mhimax{{\mhi^{\rm max}}}
\def\anti{\overline}
\def\caln{{\cal N}}
\def\calm{{\cal M}}
\def\del{\delta}
\def\half{{1\over 2}}
\def\hi{\h_1}
\def\hii{\h_2}
\def\hiii{\h_3}
\def\mhi{m_{\hi}}
\def\mhii{m_{\hii}}
\def\mhiii{m_{\hiii}}
\def\mpp{m_{PP}}
\def\mzlam{m_{Z\lam}^2}

\def\tauptaum{\tau^+\tau^-}

\def\zstar{Z^{\star}}

\def\rts{\sqrt s}

\def\h{h}
\def\mh{m_{\h}}
\def\a{a}
\def\ma{m_{\a}}

\def\lam{\lambda}

\def\eg{{\it e.g.}}

\def\taup{\tau^+}
\def\taum{\tau^-}

\def\lsim{\mathrel{\raise.3ex\hbox{$<$\kern-.75em\lower1ex\hbox{$\sim$}}}}
\def\gsim{\mathrel{\raise.3ex\hbox{$>$\kern-.75em\lower1ex\hbox{$\sim$}}}}
\def\@versim#1#2{\vcenter{\offinterlineskip
        \ialign{$\m@th#1\hfil##\hfil$\crcr#2\crcr\sim\crcr } }}
\def\zstar{Z^\star}

\def\ie{{\it i.e.}}

\def\gam{\gamma}

\def\anti{\overline}
\def\pbi{~{\rm pb}^{-1}}
\def\fbi{~{\rm fb}^{-1}}

\def\gev{\,{\rm GeV}}

\def\hsm{h_{SM}}

\def\hl{h^0}
\def\hh{H^0}
\def\ha{A^0}
\def\hp{H^+}

\def\hpm{H^{\pm}}
\def\mhl{m_{\hl}}

\def\mha{m_{\ha}}
\def\mhp{m_{\hp}}

\def\tanb{\tan\beta}
\def\cotb{\cot\beta}
\def\mt{m_t}
\def\mb{m_b}
\def\mz{m_Z}
\def\mw{m_W}

\def\wp{W^+}

\def\h{h}
\def\mh{m_{\h}}

\documentclass[smus]{snow2e}
\input psfig.sty
\begin{document}
\title{
{\large
             \hspace*{\fill}\phantom{help}
             \hspace*{\fill}\phantom{help}
             \hspace*{\fill} UCD-96-26  \\
             \hspace*{\fill} SCIPP-96/47 \\
             \hspace*{\fill} LBNL-39469 \\
             \hspace*{\fill} October, 1996 \\
             \hspace*{\fill}  \\
             \hspace*{\fill}  \\
             \hspace*{\fill}  \\
}
Will at least one of the Higgs bosons of the next-to-minimal
supersymmetric extension of the Standard Model be
observable at LEP2 or the LHC?\thanks{
To appear in ``Proceedings of the 1996 DPF/DPB Summer Study
on New Directions for High Energy Physics''.
Work supported in part by the Department of Energy
and in part by the Davis Institute for High Energy Physics.}}

\author{
John F. Gunion\\ 
{\it Davis Institute for High Energy Physics, University
of California, Davis, California 95616} \and
Howard E. Haber \\ 
{\it Santa Cruz Institute for Particle Physics, University of California,
Santa Cruz, California 95064} \and
Takeo Moroi \\ 
{\it Lawrence Berkeley National Laboratory, Berkeley, California 94720}
}

\maketitle

\thispagestyle{empty}\pagestyle{plain}

\begin{abstract}
We demonstrate that there are regions of parameter space
in the next-to-minimal (\ie\ two-Higgs-doublet, one-Higgs-singlet
superfield) supersymmetric extension of the SM
for which none of the Higgs bosons are observable either at LEP2
with $\rts=192\gev$ and an integrated luminosity of 
$L=1000\pbi$ or at the LHC with $L=600\fbi$.
\end{abstract}

\section{Introduction}

It has been demonstrated that detection of at least one of the
Higgs bosons of the minimal supersymmetric standard model (MSSM)
is possible either at LEP2 or at the LHC
throughout all of the standard $(\mha,\tanb)$ parameter space
(for a recent review, see Ref.~\cite{dpfreport}).
Here, we reconsider this issue in the context of
the next-to-minimal supersymmetric standard model (NMSSM) \cite{eghrz}
in which there is one Higgs singlet superfield
in addition to the two Higgs doublet superfields of the MSSM.
(The NMSSM Higgs sector is taken to be CP-conserving.)
We will demonstrate that there are regions of parameter space
for which none of the NMSSM Higgs bosons can be detected at
either LEP2 or the LHC.
This result should be contrasted with the NLC no-lose
theorem \cite{nlcnolose}, according to which at least one of the CP-even Higgs
bosons\footnote{We
use the generic notation $\h$ ($\a$) for a CP-even (CP-odd) Higgs boson.}
of the NMSSM will be observable in the $\zstar\to Z\h$
production mode. However, we do find that
the parameter regions for which Higgs boson
observability is not possible at LEP2 or the LHC represent a small
percentage of the total possible parameter space.

Many detection modes are involved in establishing the LHC no-lose theorem
for the MSSM. A more than adequate set is:
1) $\zstar\to Z\h$ at LEP2; 2) $\zstar\to \h\a$ at LEP2;
3) $gg\to \h\to\gam\gam$ at LHC; 4) $gg\to\h\to Z\zstar~{\rm or}~ZZ\to 4\ell$
at LHC; 5) $t\to\hp b$ at LHC;
6) $gg\to b\anti b \h,b\anti b\a \to b\anti b \tauptaum$ at LHC;
7)  $gg\to\h,\a\to\tauptaum$ at LHC.
Additional LHC modes that have been considered include:
a) $\a\to Z\h$; b) $\h\to\a\a$;
c) $\h_j\to\h_i\h_i$; d) $\a,\h\to t\anti t$.  Because
of the more complicated Higgs self interactions,
b) and c) cannot be reliably computed in the NMSSM without
additional assumptions. The Higgs mass values for
which mode a) is kinematically allowed can be quite different
than those relevant to the MSSM and thus there are
uncertainties in translating ATLAS and CMS results for the MSSM
into the present more general context. Finally, mode d) is currently
of very uncertain status and might turn out to be either more
effective or less effective than current estimates.
Thus, to be conservative, we excluded from our considerations any choice
of NMSSM parameters for which the modes a)-d) might be relevant.
Even over this restricted region of parameter space, we shall
demonstrate that NMSSM parameter choices can be found such that there are
no observable Higgs signatures at either LEP2 or the LHC.

\section{Parameters and Scanning Procedure}

In order to specify a point in NMSSM parameter space, we have adopted
the following procedure.
\begin{itemize}
\item Employ a basis in which only the first neutral Higgs field
has a vev: $\langle \phi_1\rangle=v=246\gev$.
In this basis, the $(11,12,21,22)$
elements of the Higgs mass-squared matrix (denoted ${\cal M}^2$ below)
take the simple form
\begin{equation}
\left(\begin{array}{cc}
\mz^2+\mzlam s_{2\beta}^2+\delta_{11} & \mzlam
s_{2\beta}c_{2\beta}+\delta_{12}
\\
\mzlam s_{2\beta}c_{2\beta}+\delta_{12} & \mpp^2-\mzlam
s_{2\beta}^2+\delta_{22}
\\ \end{array}\right)
\end{equation}
where $\lam$ appears in the superpotential
in the term $W\ni \lam \hat H_1 \hat H_2 \hat N$,
$\mzlam\equiv\half\lam^2 v^2-\mz^2$, and $\del_{11,12,22}$
are the radiative corrections\footnote{These
have been computed following the procedures of Ref.~\cite{habertwoloop}.}
(which are independent of $\lam$ and $\mpp$,
but depend on $\tanb$ and $\mt$ --- we take $\mt=175\gev$).
We note that there are enough parameters
in the NMSSM model superpotential and soft-supersymmetry-breaking
terms that the $\calm_{13,23,33}^2$ entries can have arbitrary
values.  (Specific Planck scale boundary conditions could restrict
these latter $\calm^2$ entries and thereby impose restrictions
on the allowed parameter space beyond those
described below; such boundary conditions will not be imposed here.)
\item Pick a value for $\tanb$ and a value for $\mhi\leq\mhi^{\rm max}$,
where $\mhi^{\rm max}=\calm_{11}(\lam=\lam_{\rm max})$.
The crucial ingredient in limiting the scan is the upper limit
of $\lam_{\rm max}=0.7$ \cite{lamlimit}
obtained by requiring that $\lam$ remain
perturbative during evolution from scale $\mz$ to the Planck
scale.
\item Pick values for the angles $-\pi/2\leq\alpha_1\leq+\pi/2$,
$0\leq\alpha_2\leq 2\pi$, and $0\leq\alpha_3\leq \pi/2$ that appear
in the matrix $V$ which diagonalizes the CP-even Higgs mass-squared
matrix via $V^\dagger {\cal M}^2 V={\rm diag}(\mhi^2,\mhii^2,\mhiii^2)$:
\begin{equation}
V= \left(\begin{array}{ccc}
c_1 & -s_1c_3 & -s_1s_3 \\
s_1c_2 & c_1c_2c_3-s_2s_3 & c_1c_2s_3+s_2c_3 \\
s_1s_2 & c_1s_2c_3+c_2s_3 & c_1s_2s_3-c_2c_3 \end{array} \right)
\end{equation}
where $c_1=\cos\alpha_1$, and so forth. It is useful to note that
\begin{eqnarray}
\mhii^2 &\leq& {[\mhi^{\rm max}]^2-V_{11}^2 \mhi^2 \over 1 - V_{11}^2}
\label{mhiilim} \\
\mhiii^2&\leq &{[\mhi^{\rm max}]^2-V_{11}^2 \mhi^2-V_{12}^2\mhii^2 \over
              1-V_{11}^2-V_{12}^2 }\,.
\label{mhiiivalue}
\end{eqnarray}
\item Pick a value $\lam_{\rm min}\leq\lam\leq \lam_{\rm max}$, and compute
\begin{eqnarray*}
&\mhii^2 = { V_{13}\calm^2_{12}-V_{23}\calm^2_{11}-\mhi^2 V_{11}
(V_{21}V_{13}-V_{23}V_{11}) \over
V_{12}(V_{22}V_{13}-V_{23}V_{12}) }\,, &\\
&\mhiii^2 = { V_{12}\calm^2_{12}-V_{22}\calm^2_{11}-\mhi^2 V_{11}
(V_{21}V_{12}-V_{22}V_{11}) \over
V_{13}(V_{23}V_{12}-V_{22}V_{13}) } \,, &\\
&\mpp^2 = \sum_{i=1,2,3} V_{2i}^2 m_{h_i}^2+\mzlam
s_{2\beta}^2-\del_{22}\,,\phantom{m1} &\\
&\mhp^2 = \mpp^2-\mzlam\,. \phantom{m  \mpp^2 = \sum m m m m m} &
\end{eqnarray*}
The lower limit on $\lam$ is given by
\begin{equation}
\lam_{\rm
min}^2 v^2=2\left[{\mhi^2-\del_{11}-\mz^2\over
s_{2\beta}}+\mz^2\right]\,,
\end{equation}
which is obtained by noting that $\mhi^2\leq \calm_{11}^2$.
If $\lam_{\rm min}^2<0$ then use $\lam_{\rm min}=0$. It is consistent
to consider only those
$\alpha_i,\lam$ values such that $\mhiii^2\geq\mhii^2\geq\mhi^2$.
Further restrictions are imposed on the $m_{\h_i}^2$ as follows.
First, we require that
$\mhiii\le 2\mhi$, in which case the decays
$\h_2\to\h_1\h_1$, $\h_3\to\h_1\h_1$ and $\h_3\to\h_2\h_2$
are all kinematically disallowed.
(If kinematically allowed, such decays are model dependent and
could be dominant; their experimental accessibility would
have to be evaluated.) Second, we require that $\mhiii\leq 2\mt$
so that the decays $\h_{1,2,3}\to t\anti t$ are forbidden.
\item
The CP-odd mass-squared matrix takes the form
\begin{equation}
\caln^2 = \left(\begin{array}{cc}
\mpp^2 & \cdot \\ \cdot & \cdot \\
\end{array}\right)\,,
\end{equation}
where the unspecified entries may take on any value given
the parameter freedom of the model.  For simplicity, we
assume that only one CP-odd scalar, the $\a$
(which must have $\ma^2\leq\mpp^2$), is possibly light and that
the other is heavy and, therefore, unobservable.  In principle,
we could scan $0\leq\ma\leq\mpp$.  However,
we impose three additional restrictions on $\ma$ as
follows.  In order to avoid the presence of the
model-dependent, possibly dominant $h_{1,2,3}\to \a\a$ decays,
we restrict the scan to $\ma\geq \mhiii/2$.  In particular, this implies
that no $\ma$ scan is possible if $\mpp\leq\mhiii/2$.  We also
impose the restrictions: $\ma\leq 2\mt$, so
that $\a\to t\anti t$ decays are forbidden; and 
$\ma\leq \mz+\mhi$, which implies that
the model-dependent decays $\a\to Z\h_{1,2,3}$ are absent.
\end{itemize}
We emphasize that there may be parameter choices,
for which no Higgs bosons of the NMSSM
are observable, that lie outside the restricted portion of parameter space
that we search. Our goal here is not to fully delineate all
problematical parameter choices, but rather to demonstrate the
existence of parameters for which
it is guaranteed that no NMSSM Higgs boson can be found
without increased LEP2 energy and/or luminosity,
or increased LHC luminosity or LHC detector improvements.

\section{Detection Modes}

In order to assess the observability of modes 1)-7) we need the
couplings of the $\h_{1,2,3}$ and $\a$.  Those required are:
\begin{eqnarray}
&& ZZ\h_i,WW\h_i: [{g\mz\over c_W},g\mw] V_{1i}\label{zzh}\\
&& Z\h_i \a: {g\over 2c_W} V_{2i}\label{zha}\\
&& t\anti t\h_i: {g\mt\over 2\mw} (V_{1i}+V_{2i}\cotb)
\label{tth} \\
&& b\anti b\h_i: {g\mb\over 2\mw} (V_{1i}-V_{2i}\tanb)
\label{bbh} \\
&& t\anti t\a,b\anti b\a: {g\mt\over 2\mw}\cotb,~~{g\mb\over 2\mw}\tanb
\label{ttbba}
\end{eqnarray}
As already noted, we do not search parameter regions in which
the very model-dependent Higgs self-couplings would be needed.

Within the domain of parameter space that we search, we
evaluate the potential of modes 1)-7) as follows.
For the LEP2 modes 1) and 2), we require 30 and 50 events, respectively,
for $L=1000\pbi$, before any cuts, branching ratios, or efficiency factors.
For the LHC modes 3)-7), we require $5\sigma$ statistical significance
for $L=600\fbi$.  The individual mode treatments are as follows.
\begin{itemize}
\item
For the $\h_i\to\gam\gam$ and $\h_i\to Z\zstar,ZZ\to 4\ell$ modes, 3) and 4),
we compute the number of events as compared to predictions for the
SM Higgs boson, and then compute the resulting statistical significance
assuming scaling proportional to the signal event rate.  The most
optimistic SM Higgs
statistical significances for the $\gam\gam$ and $4\ell$ channels
as a function of Higgs mass are those from CMS \cite{CMSTN},
Fig.~4 ($\gam\gam$) and Fig.~8 ($Z\zstar$),
and Tables 35 and 36 ($ZZ$ ) of Ref.~\cite{ATLASnote}.
We increase these $L=100\fbi$ statistical significances
by a factor of $\sqrt 6$ for $L=600\fbi$
and then apply the NMSSM corrections.
\item
For the $t\to \hp b$ detection
mode 5) we employ the $L=600\fbi$ contours, Fig.~76,
of Ref.~\cite{latestplots}.
We note that when $t\to \hp b$ is kinematically allowed,
the $\hp\to \wp \h_{1,2,3}$ decays are forbidden for the $\mhi$
values we consider here. Thus, the $\hp$ decays are exactly
as in the MSSM and the MSSM results can be employed `as is'
when the $5\sigma$ contour is specified as a function of $\mhp$
and $\tanb$.
\item
For the $b\anti b \h$ and $b\anti b\a$ final states we
refer to the $L=100\fbi$ statistical significances
quoted for the MSSM model $b\anti b\ha$ process
at $\tanb=10$ in Table~34 of Ref.~\cite{latestplots}
and the input $\br(\ha\to \tauptaum)$ from Fig.~22 ($\tanb=10$ results)
of Ref.~\cite{latestplots}. From these results we compute a standard
statistical significance for $\tanb=1$, $\br(\a\to\tauptaum)=1$,
and $L=600\fbi$. Statistical significances in the NMSSM model
are obtained for the $\h_i$ and $\a$ by multiplying these standard
statistical significances by the appropriate $(b\anti b\h_i)^2$
enhancement factor or by $(b\anti b\a)^2=\tan^2\beta$ and by
the computed $\tauptaum$ branching ratio of the Higgs boson in question.
Recall that we do not search parameter regions
for which the $\tauptaum$ branching ratios would be uncertain
due to Higgs pair decay channels being kinematically allowed.
\item
Finally, we assume that mode 7) is only relevant for the $\a$
(as in the MSSM). However, we cannot directly
use the discovery region shown for $L=300\fbi$ in Fig.~53 of
Ref.~\cite{latestplots} since $\ha\to Z\hl$ decays deplete
the $\tauptaum$ branching ratio for $\mha\gsim 190\gev$. Thus, we
use an optimistic limit for this mode's $L=600\fbi$ region of
viability; $\geq 5\sigma$ is assumed to be
achieved in this mode for $\tanb\leq 4$ if $100\leq\ma\leq 350\gev$.
\end{itemize}
If none of the Higgs bosons $\h_{1,2,3}$, $\a$ or $\hpm$ are observable
as defined above
we declare a parameter point in our search to be a ``point of
unobservability'' or a ``bad point''.

\section{Results}

We now summarize our results.  We find that if $\tanb\lsim 1.5$
then all parameter points that are included in our search are observable
for $\mhi$ values up to the maximum allowed ($\mhi^{\rm max}\sim 137\gev$
for $\lam_{\rm max}=0.7$, after including radiative corrections).
For such low $\tanb$, the LHC $\gam\gam$ and $4\ell$ modes allow
detection if LEP2 does not. For high $\tanb\gsim 10$, the
parameter regions where points of unobservability are found
are also of very limited extent, disappearing as the $b\anti b\h_{1,2,3}$
and/or $b\anti b\a$ LHC modes allow detection where LEP2 does not.
However, significant portions of
searched parameter space contain points of unobservability
for moderate $\tanb$ values. That such $\tanb$ values
should be the most `dangerous'
can be anticipated from the MSSM results.  It is well-known
(see, for example, Ref.~\cite{dpfreport}) that there is a wedge
of MSSM parameter space at moderate $\tanb$ and with $\hh$ and $\ha$
masses above about $200\gev$ for which the only observable
MSSM Higgs boson is the light SM-like $\hl$, and that it can
only be seen in the $\gam\gam$ mode at the LHC
($\mhl+\mz,\mhl+\mha > \rts$ at LEP2).  By choosing $\mhi$ and $\ma$
in the NMSSM
so that $\mhi+\mz$ and $\mhi+\ma$ are close to or above the $\rts$ of LEP2,
then, by analogy, at moderate $\tanb$ 
we would need to rely on the $\h_{1,2,3}\to \gam\gam$ modes.
However, in the NMSSM, parameter choices are possible for which all the
$WW\h_{1,2,3}$ couplings are reduced relative to SM strength.
This reduction will
suppress the $\gam\gam$ couplings coming from the $W$-boson
loop. All the $\h_i\to\gam\gam$ widths
can be sufficiently smaller than the somewhat enhanced $b\anti b$ widths
so that the $\gam\gam$ branching ratios are {\it all} no longer of
useful size.

\begin{figure}[htb]
\leavevmode
\begin{center}
\centerline{\psfig{file=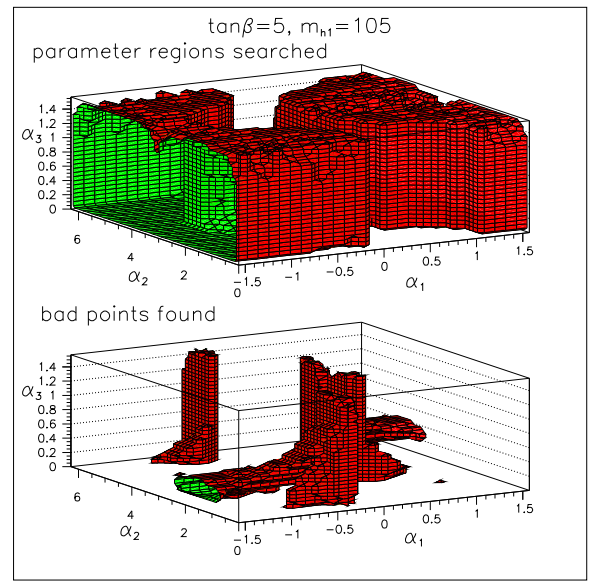,width=3.0in}}
\end{center}
\caption{For $\tanb=5$ and $\mhi=105\gev$, we display in three dimensional
$(\alpha_1,\alpha_2,\alpha_3)$ parameter space the parameter regions
searched (which lie within the surfaces shown), and the
regions therein for which the remaining model parameters can
be chosen so that no Higgs boson is observable
(interior to the surfaces shown).}
\label{tanb5}
\end{figure}

To illustrate, we shall discuss results for $\tanb=3$, $\tanb=5$ and
$\tanb=10$ (for which $\mhi^{\rm max}\sim 124\gev$, $118\gev$ and $114\gev$,
respectively) and $\mhi=105\gev$.
\begin{itemize}
\item
In Fig.~\ref{tanb5}, we display for $\tanb=5$ both the
portions of $(\alpha_1,\alpha_2,\alpha_3)$
parameter space that satisfy our search restrictions,
and the regions (termed ``regions of unobservability'')
within the searched parameter space such that,
for {\it some} choice of the remaining parameters ($\lam$ and $\ma$),
no Higgs boson will be detected
using any of the techniques discussed earlier.
\footnote{For a given $\alpha_{1,2,3}$ value
such that there is a choice of $\lam$ and $\ma$ for which no Higgs
boson is observable, there are generally
other choices of $\lam$ and $\ma$ for which at least one
Higgs boson {\it is} observable.} Relatively large regions of unobservability
within the searched parameter space are present.
\item
At $\tanb=3$, a similar picture emerges. The search region that satisfies our
criteria is nearly the same; the regions of unobservability lie mostly within
those found for $\tanb=5$, and are about 50\% smaller.
\item
For $\tanb=10$, the regions of unobservability comprise only a
very small portion of those found for $\tanb=5$.
This reduction is due to the increased $b\anti b$ couplings
of the $\h_i$ and $\a$, which imply increased $b\anti b\h_i,b\anti b\a$
production cross sections. As these cross sections become large, detection
of at least one of the $\h_i,\a$ in the
$b\anti b\taup\taum$ final state becomes
increasingly difficult to avoid. For values of $\tanb\gsim 10$,
\footnote{The precise value of the critical lower bound on $\tanb$
depends sensitively on $\mhi$.}
we find that one or
more of the $\h_i,\a$ should be observable regardless of location in
$(\alpha_1,\alpha_2,\alpha_3,\lam,\ma)$ parameter space (within
the somewhat restricted search region that we explore).
\end{itemize}

\begin{figure}[htb]
\leavevmode
\begin{center}
\centerline{\psfig{file=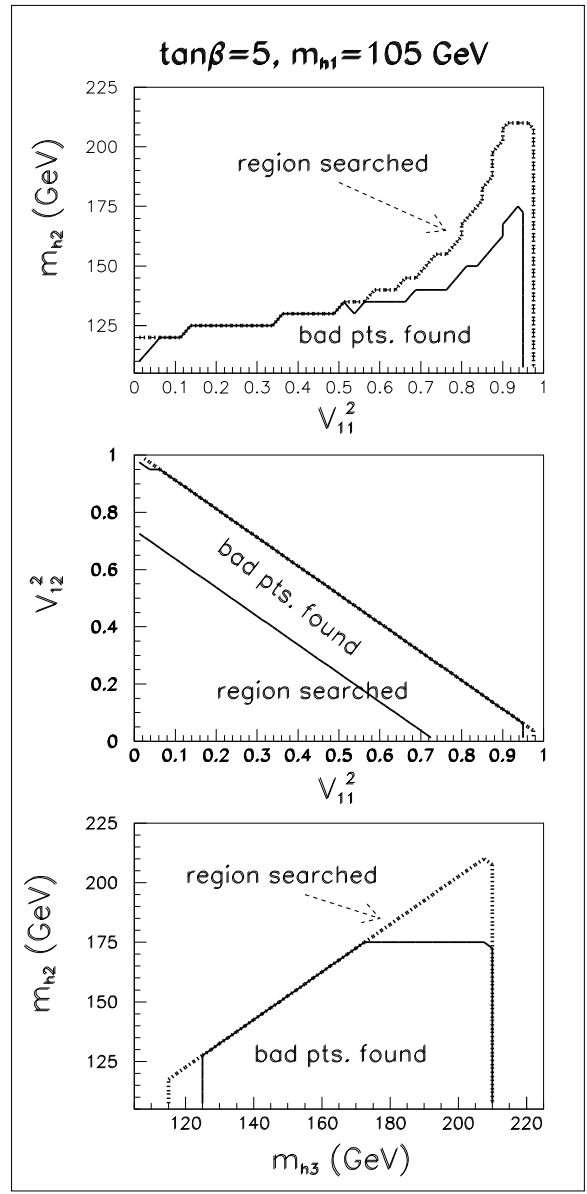,width=3.0in}}
\end{center}
\caption{For $\tanb=5$ and $\mhi=105\gev$, we display
the regions of the $(V_{11}^2,\mhii)$, $(V_{11}^2,V_{12}^2)$
and $(\mhiii,\mhii)$ parameter spaces that were searched
and the regions
therein (labeled ``bad points found'')
for which there is {\it some choice} for the
remaining NMSSM parameters such that
no Higgs boson is observable.
}
\label{v11mh2mh3}
\end{figure}

Another perspective on the parameter space and the location of
points of unobservability is provided in Fig.~\ref{v11mh2mh3}.
There, we display for $\tanb=5$ and $\mhi=105\gev$
the regions searched in the $(V_{11}^2,\mhii)$, $(V_{11}^2,V_{12}^2)$
and $(\mhiii,\mhii)$ parameter spaces, and the portion thereof
in which the remaining model parameters can be chosen
such that no Higgs boson is observable.
The $(V_{11}^2,\mhii)$ plot shows that Higgs boson unobservability is
possible for any value of $V_{11}^2$ and for all values of $\mhii$
up to the bound of Eq.~(\ref{mhiilim}), so long as $V_{11}^2\lsim 0.5$.
For $V_{11}^2\gsim 0.5$, the region of $\mhii$ for which Higgs boson
unobservability is possible does not include the highest $\mhii$ values.
The $(V_{11}^2,V_{12}^2)$ plot
shows that unobservability is possible only if $V_{11}^2+V_{12}^2\gsim 0.7$,
\ie\ the $ZZ\h_3$ coupling is reduced relative to SM strength
by $V_{13}^2\lsim 0.3$, implying that
$\h_3$ is difficult to detect in the $ZZ\to 4\ell$ mode.
The $(\mhii,\mhiii)$ plot shows that unobservability is possible
for almost all $\mhiii$ values so long as $\mhii\lsim
2\mz$. For $\mhii\lsim 2\mz$, the $\h_2$
must be detected in the relatively weak $\h_2\to Z\zstar~{\rm or}~\gam\gam$
modes; both are typically somewhat suppressed at moderate (or large) $\tanb$
by a $gg\h_2$ coupling  that is smaller than SM-strength and by an
enhanced $b\anti b$ decay width that diminishes the $Z\zstar,\gam\gam$
branching fractions.
Throughout the regions displayed in Fig.~\ref{v11mh2mh3}
where choices for the remaining model parameters can make observation
of any of the Higgs bosons impossible,
there are other choices for the remaining parameters such
that at least one Higgs boson {\it is} observable.

The mass $\mhi=105\gev$ is typical of the `intermediate'
values that yield the largest regions of unobservability.
If $\mhi\lsim 85\gev$, then
discovery of one of the $\h_i$ at LEP2 is almost certain.
As $\mhi\to \mhi^{\rm max}$,
then discovery of at least one Higgs boson at the LHC 
becomes possible over most
of parameter space, as we now describe.
As $\mhi\to \mhi^{\rm max}$,
$V_{13}^2\to 0$.~\footnote{If $V_{13}\neq 0$, then Eqs.~(\ref{mhiilim}) and
(\ref{mhiiivalue}) imply that $\mhiii\to \mhii\sim\mhi$ as $\mhi\to \mhimax$.
In this limit we have $\calm_{12}^2=\sum_{i=1,2,3} V_{1i}V_{2i}m_{\h_i}^2
\to \mhi^2 \sum_{i=1,2,3}V_{1i}V_{2i}=0$ by orthogonality of $V$.
Unless $\calm_{12}^2=0$, there is an inconsistency which can only
be avoided by simultaneously taking $V_{13}^2\to 0$.}
Since $V_{13}=-s_1s_3$, this means either
$\alpha_1\sim 0$ or $\alpha_3\sim 0$. However,
only if $\alpha_3\sim 0$ can all the Higgs bosons
be unobservable.  If $\alpha_3$ is not near 0,
$\alpha_1$ must be, in which case $V_{21}\sim 0$ and $V_{11}\sim 1$
and the $\h_1$ has completely SM-like couplings [see
Eqs.~(\ref{zzh})-(\ref{ttbba})], and for $\mhi\sim\mhi^{\rm max}$ ($\sim
118\gev$ at $\tanb=5$) $\h_1$
will be detectable in the $\gam\gam$ final state. If $\alpha_3\sim 0$,
then any value of $\alpha_1$ is possible, but (again) $\alpha_1\sim 0$
would make $\h_1$ SM-like and observable; in addition,
$\alpha_1\sim \pm \pi/2$
(\ie\ $s_1\sim \pm 1,c_1\sim 0$) yields $V_{22}\sim 0$ and $|V_{12}|\sim 1$
implying that $\h_2$ would be SM-like and observable
(in the $\gam\gam$ or $Z\zstar,ZZ$ modes). 
Thus, the only `dangerous' region is $\alpha_3\sim 0$ and
$\alpha_1\neq0,\pm\pi/2$, for which, Eq.~(\ref{mhiilim}) implies 
$\mhii\sim\mhi$ so that both
$\hii$ and $\hi$ would have to be found in the $\gam\gam$
mode.\footnote{Note that in the $\gam\gam$ channel,
the resolution is such that
extreme degeneracy, $\Delta \mh\lsim 1\gev$, is required before
we must combine signals.}
If the value of $\alpha_2$ is such
that neither $s_2$ nor $c_2$ is small, then both $V_{21}$ and $V_{22}$
can be substantial, and the $\gam\gam$
mode can be suppressed for both $\h=\hi$ and $\h=\hii$
by a combination of $t\anti t\h$ coupling
suppression (to diminish $gg\to\h$ production) and $b\anti b\h$ coupling
enhancement (as natural for moderate or large $\tanb$). The latter
enhances the $b\anti b$ partial width and diminishes the $\h\to\gam\gam$
branching ratio. The moderate $\tanb\sim 5$ value
makes it possible to have the required $b\anti b\h$ coupling
enhancement without it being so large as to make the $\h\to\tauptaum$ mode
observable in $b\anti b\h$ production.

It is useful to present details on what goes wrong at a typical
point of unobservability. For $\tanb=5$ and $\mhi=105\gev$,
no Higgs boson can be observed for $\ma=103\gev$ if
$\alpha_1=-0.479$, $\alpha_2=0.911$, $\alpha_3=0.165$, and $\lam=0.294$
(for which $\mhii=124\gev$, $\mhiii=206\gev$, $\mhp=201\gev$,
and $\mpp=186\gev$). The corresponding $V$ matrix entries are:
\begin{equation}
V=\left( \begin{array}{ccc}
    0.887 &  0.455 &  0.0757 \\
  -0.283  &  0.407 &  0.869 \\
  -0.364  &  0.792 & -0.490 \end{array} \right)\,.
\end{equation}
{}From the $V_{ij}$, and the value of $\tanb$, we compute (relative to
the SM values)
\begin{eqnarray*}
(VV\h_1)^2= 0.79 & (VV\h_2)^2=0.21 & (VV\h_3)^2=0.006 \\
(b\anti b \h_1)^2=5.3 & (b\anti b \h_2)^2=2.5 & (b\anti b\h_3)^2=18 \\
(t\anti t\h_1)^2=0.69 & (t\anti t \h_2)^2=0.29 & (t\anti t\h_3)^2=0.062
\end{eqnarray*}
where $V=W$ or $Z$.
Note that $\h_3$ has very small couplings to $VV$.

The manner in which this point
escapes discovery is now apparent. First,
the minimum values required for the $(b\anti b\h_i)^2$ values for $\h_i$
observability in the $\tauptaum$ mode are:  53 ($i=1$); 32 ($i=2$); 35 ($i=3$).
The actual values all lie below those required.
Observation of the $\a$ at $\ma=103\gev$ (without adding in
the much smaller overlapping $\h_1$ signal)
would require $\tanb=8$.
Regarding the other discovery modes,
$\h_1$ and $\h_2$ are both in the mass range for which the $\gam\gam$
mode is potentially viable and the $\h_3$
is potentially detectable in the $ZZ\to 4\ell$ channel.
However, the suppressed $t\anti t\h_{1,2,3}$
couplings imply smallish $gg$ production rates for $\h_{1,2,3}$.
Relative to a SM Higgs of the same mass we have:
\begin{equation}
{(gg\h_i)^2\over (gg\hsm)^2}=0.58~(i=1);~~~0.43~(i=2);~~~0.15~(i=3)\,.
\end{equation}
(Note that these strengths are not simply the $(t\anti t\h_i)^2$
magnitudes due to enhanced $b$-quark loop contributions which
interfere with the $t$-quark loop contributions at amplitude level.)
Further, the enhanced Higgs decay rate to $b\bar b$ and the reduced
$W$-loop contributions to the $\gam\gam$ coupling suppress the
$\gam\gam$
branching ratios of $\h_1$ and $\h_2$ relative to SM expectations.  We find:
\begin{equation}
{\br(\h_i\to \gam\gam) \over \br(\hsm\to\gam\gam)}
=0.18~(i=1)\,;~~~0.097~(i=2)\,;
\end{equation}
\ie\ suppression sufficient to make $\h_1$ and $\h_2$ invisible
in the $\gam\gam$ mode. The suppressed $ZZ\h_3$ coupling
and the enhanced $\h_3\to b\bar b$ decays
are sufficient to suppress $\br(\h_3\to ZZ)$
much below SM expectations:
\begin{equation}
{\br(\h_3\to ZZ)\over \br(\hsm\to ZZ)}=0.11\,,
\end{equation}
\ie\ such that the $4\ell$ signal has a significance of only $1.5\sigma$,
even though a SM Higgs of this mass would yield a $\sim 37\sigma$ signal.

In short, there is enough flexibility due to the addition of the singlet
Higgs field (which has no couplings to SM fermions and vector bosons!)
for {\it all} the Higgs bosons to escape
detection for certain choices of model parameters,  provided $\tanb$
is moderate in size. Moderate $\tanb$ implies
that $\h\to\gam\gam$ decays for light Higgs are
suppressed, while at the same time
$b\anti b \h$ production is not adequately enhanced
for detection of the $\h\to \tauptaum$ mode.

\section{Discussion and Conclusions}

The regions of NMSSM parameter space where no Higgs boson
can be detected will expand if
full $L=600\fbi$ ($L=1000\pbi$) luminosity is not available at the LHC
(LEP2) or efficiencies are smaller than anticipated.
Conversely, these ``regions of unobservability''
could decrease substantially (perhaps disappear)
with improved efficiency (\eg\ due to the
expanded calorimeter option discussed in Ref.~\cite{latestplots})
in the $\tau\tau$ final state or higher luminosity.
These issues will be pursued elsewhere.

We have explicitly neglected supersymmetric (SUSY)
decay modes of the Higgs bosons
in our treatment.  If these decays are important,
the regions of unobservability found without using the SUSY final states will
increase in size.  However, Higgs masses in the
regions of unobservability
are typically modest in size ($100-200\gev$), and
as SUSY mass limits increase with LEP2 running this additional
concern will become less relevant.  Of course, if SUSY
decays are significant, detection of the Higgs
bosons in the SUSY modes might be possible, in which case
the regions of unobservability might decrease in size. Assessment of this
issue is dependent upon a specific model for soft SUSY breaking
and will not be pursued here.

Finally, although we cannot establish a no-lose
theorem for the NMSSM Higgs bosons at LEP2 and the LHC
(in contrast to the no-lose theorems applicable to
the NLC Higgs search with $\rts\gsim 300\gev$), the regions of complete
Higgs boson
unobservability appear to constitute a small fraction of the
total model parameter space.
It would be interesting to see whether or not these
regions of unobservability correspond to unnatural choices
for the Planck scale supersymmetry-breaking parameters.

%

\end{document}